\documentclass[preprint,superscriptaddress,tightenlines,nofootinbib,amsmath,amssymb,aps]{revtex4}
\usepackage[T1]{fontenc}
\usepackage{graphicx}
\usepackage{dcolumn}
\usepackage{bm}
\usepackage{xcolor}
\usepackage{tabularx}
\usepackage{caption}

\newcommand{\beq}{\begin{equation}}
\newcommand{\eeq}{\end{equation}}
\newcommand{\beqn}{\begin{eqnarray}}
\newcommand{\eeqn}{\end{eqnarray}}

\newcommand\barparen[1]{\overset{{\scriptscriptstyle\text{~(---)}}}{#1}}

\newcommand{\mevc}{\ensuremath{\,{\mathrm{MeV}/c^2}}}

\newcommand{\ee}{\ensuremath{e^+e^-}}
\newcommand{\Br}{\ensuremath{\mathcal{B}}}

\newcommand{\al}{\ensuremath{a}}
\newcommand{\be}{\ensuremath{b}}
\newcommand{\figa}{{\bf a}}
\newcommand{\figb}{{\bf b}}

\newcommand{\lep}{\ensuremath{\ell}}

\newcommand{\Dn}{\ensuremath{D^0}}
\newcommand{\Db}{\ensuremath{\bar{D}{}^0}}
\newcommand{\ddb}{\Dn-\Db}

\newcommand{\Kn}{\ensuremath{K^0}}
\newcommand{\Kb}{\ensuremath{\bar{K}^0}}

\newcommand{\Ks}{\ensuremath{K^0_S}}
\newcommand{\Kl}{\ensuremath{K^0_L}}
\newcommand{\pp}{\ensuremath{\pi^+ \pi^-}}
\newcommand{\pin}{\ensuremath{\pi^0}}

\newcommand{\Kba}{\ensuremath{|\Kb\rangle}}
\newcommand{\Kna}{\ensuremath{|\Kn\rangle}}

\newcommand{\Dna}{\ensuremath{{|\Dn\rangle}}}
\newcommand{\Dba}{\ensuremath{|\Db\rangle}}

\newcommand{\qpd}{\ensuremath{\left( \frac{q}{p} \right)_{\!\!D}}}
\newcommand{\pqd}{\ensuremath{\left( \frac{p}{q} \right)_{\!\!D}}}
\newcommand{\qpk}{\ensuremath{\left( \frac{q}{p} \right)_{\!\!K}}}
\newcommand{\pqk}{\ensuremath{\left( \frac{p}{q} \right)_{\!\!K}}}

\newcommand{\rd}{\ensuremath{r_D}}
\newcommand{\rdz}{\ensuremath{r_D^{0}}}
\newcommand{\rdp}{\ensuremath{r_D^{-}}}
\newcommand{\ord}{\ensuremath{\overline{r}_D}}
\newcommand{\dkp}{\ensuremath{\delta_{K\pi}}}
\newcommand{\dzz}{\ensuremath{\delta^{0}}}
\newcommand{\dpm}{\ensuremath{\delta^{-}}}
\newcommand{\ctau}{\ensuremath{c}-\ensuremath{\tau}}

\newcommand{\Epl}{\ensuremath{E_{\pi \lep}}}
\newcommand{\Eps}{\ensuremath{E_{\pi \lep}^2}}
\newcommand{\Ppl}{\ensuremath{{\bf p}_{\pi \lep}}}
\newcommand{\Pps}{\ensuremath{p_{\pi \lep}^2}}

\begin{document}

\title{\boldmath Measurement of \ddb\ mixing parameters using semileptonic 
decays of neutral kaon}
\author{P. Pakhlov,}
\author{V. Popov}
\affiliation{P.N. Lebedev Physical Institute of the RAS, Moscow, Russia}
\email{popovve@lebedev.ru}
\date{\today}

\begin{abstract}
We propose a new method to extract \ddb\ mixing parameters using 
the $\Dn \to \Kb \pin$ decay with the \Kb\ reconstructed in the 
semileptonic mode. Although a $\Kn \to \pi^\pm \lep^\mp \nu_\ell$ decay 
suffers from low statistics and complexity of the secondary vertex 
reconstruction in comparison to the standard $\Ks \to \pp$ vertex, 
it provides much richer, sometimes unique information about the initial state 
of a \Kn-meson produced in a heavy-flavor hadron decay. In this paper 
it is shown that the reconstruction of the chain $\Dn \to \Kn (\pi^\pm
\lep^\mp \nu_\lep) \pin$ allows one to extract the strong phase 
difference between the doubly Cabibbo-suppressed and Cabibbo-favored 
decay amplitudes, which is of key importance for determination of the 
\ddb-mixing parameters.
\end{abstract}

\maketitle

\section{Introduction}\label{sec:intro}

Neutral meson oscillations and CP violation provide one of the most powerful 
probes in searching for New Physics (NP), setting bounds on flavor structure 
of NP at the TeV scale and above and severely constraining possible 
extensions of the Standard Model (SM)~\cite{Isidori:2010kg}. While
oscillations in the $K^0$ and $B^0_{(s)}$ systems have been well studied 
for a long time, the \ddb\ mixing was established in 2007 
only~\cite{Aubert:2007wf,Staric:2007dt}. The \Dn\ system is unique in 
circulating internal $d$-type quarks in the underlying mixing box diagram 
and thus provides complimentary information on possible NP effects. 
The recent observation of the direct CP violation in charm sector by LHCb
attracts even more interest in this field~\cite{Aaij:2019kcg}.

From the theory side, the disadvantage of the \ddb\ mixing is caused by 
large long-distance effects. Some progress in calculations of the mixing 
parameters can be expected from quickly progressing lattice 
computations~\cite{Chang:2017zaw,Simone:2016bbg,Bazavov:2017weg}. Thus, 
one can hope that there will be reliable predictions with which experimental
measurements can be compared.

The largest sensitivity in mixing measurements has been achieved from the 
measurements of the time-dependent decay rates in the $\Dn \to K^+ \pi^-$ 
process, referred to as "wrong-sign" decay. However, in spite of the 
unprecedental high significance of the \ddb\ mixing observation and precise 
measurement of its strength~\cite{Ko:2014qvu, CIBINETTO:2014jsa, Aaij:2016roz}, 
the fundamental mixing parameters still need to be extracted from these 
measurements and compared to the SM predictions. It is sufficient to mention 
that the parameter $x$ remains consistent with 0 within a $2\sigma$ 
uncertainty~\cite{Amhis:2016xyh}. The main problem arises from the unknown 
strong-phase difference, $\delta$, between doubly Cabibbo-suppressed (DCS) 
and Cabibbo-favored (CF) decays, that rotates the measured mixing parameters 
relative to their true values. 

In this paper we discuss a possibility of using of $\Dn \to \Kn \pin$ decays 
with $\Kn$ reconstructed in the semileptonic mode to measure $\delta$. 
While the semileptonic mode, $\Kn \to \pi^\pm \ell^\mp \nu$, is rather rare for 
\Ks, it can be more informative relative to the standard $\Ks \to \pi \pi$ 
decay mode. The proposed method can be applied in the already running 
LHCb~\cite{Bediaga:2018lhg} and Belle II~\cite{Abe:2010gxa} experiments as
well as at the future Super \ctau\ factory~\cite{Bondar:2013cja, Chen:1997at}, 
where the expected high integrated luminosity compensates the smallness of  
$\Br(\Ks \to \pi^\pm \ell^\mp \nu)$.

\begin{figure}[htb]
\centering
\includegraphics[width=0.95\linewidth]{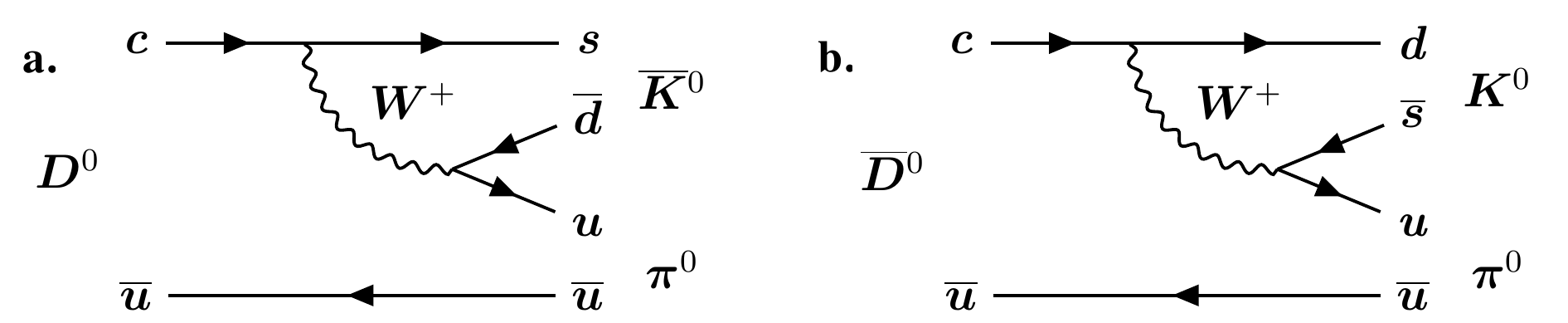}
\caption{Feynman diagrams for \figa\ CF and 
\figb\ DCS $\Dn \to \barparen{K}{}^0 \pi^0$ decays.}
\label{fig:dia}
\end{figure}

The key idea of this paper is that \Dn\ decays to \Kba\ via CF and \Kna\ via 
DCS mechanisms (Fig.~\ref{fig:dia}), thus producing $\al \Kna + \be \Kba$ 
state, where a relative complex phase between the \al\ and \be\ is the 
strong phase difference between $\Kb \pi^0$ and $\Kn \pi^0$, which is equal 
to the sought $\delta$ due to isospin symmetry. The evolution of the 
$\al \Kna + \be \Kba$ state to pure \Kna\ or \Kba\ (that are fixed at the 
time of semileptonic \Kn\ decay by the lepton sign) provides a more powerful 
tool to measure \al\ and \be\ as will be shown below. 

\section{Mixing parameters in the $\mathbf{D^0}$ system}
\label{sec:dd}

We briefly remind the basic mixing formalism. Solving the Schr\"odinger 
equation in the effective Hamiltonian approach one could obtain the following 
evolution equations for physical states produced as pure $\Dn (\Db)$: 
\beqn
\label{eq:devol}
|\Dn_{phys}(t)\rangle = g_+(t) \Dna - \qpd g_-(t) \Dba, \nonumber \\
|\Db_{phys}(t)\rangle= g_+(t) \Dba - \pqd g_-(t) \Dna ,
\eeqn
where $g_\pm = \frac{1}{2} \left( e^{-i \lambda_2 t} \pm e^{-i \lambda_1 t} \right)$ 
for eigenvalues $\lambda_{1,2} = m_{1,2} - \frac{i}{2} \Gamma_{1,2}$. It is 
conventional to describe evolution with dimensionless mixing parameters 
\textit{x, y}:
\beq
x \equiv \frac{\Delta m}{\Gamma}, \qquad \qquad 
y \equiv \frac{\Delta \Gamma}{2\Gamma},
\eeq
where $\Gamma=(\Gamma_{1}+\Gamma_{2})/2$. 

The main problem in \textit{x, y} determination arises because of the 
contribution of DCS decay in addition to the mixing to the "wrong sign" 
final state. While the mixing contribution can be disentangled from the DCS 
one by \Dn\ decay time study~\cite{Bergmann:2000id}:
\beqn
\hspace*{-0.4cm} R^+ \!(t)\!=\! \left( \! \rd \! + \! \left| \qpd\right| \sqrt{\rd} \Gamma t (y'\cos\phi_D \! - \! x' \sin\phi_D) \! + \! \left| \qpd\right|^2 
\frac{(\Gamma t)^2}{4} (x^2 \! + \! y^2) \!
\right) \! e^{-\Gamma t} \! , \nonumber \\
\hspace*{-0.4cm} R^- \! (t)\!= \!\left( \! \ord \! + \! \left| \pqd \right| \sqrt{\ord} \Gamma t (y'\cos\phi_D \! + \! x' \sin\phi_D) \! + \! 
\left|\pqd \right|^2 \frac{(\Gamma t)^2}{4} (x^2 \! + \! y^2) \!
\right) \! e^{-\Gamma t} 
\! ,
\label{eq:tdepend}
\eeqn
the measured parameters remain biased, as in the measured time-dependent 
decay rates \textit{x, y} enter as linear combinations,
\beqn
x' & = & x \cos \dkp + y \sin \dkp, \nonumber \\
y' & = & y \cos \dkp - x \sin \dkp,
\eeqn
where \dkp\ is a strong phase difference between the CF and DCS $D$ decays. 
In Eq.~\ref{eq:tdepend} $\phi_D$ is the mixing weak phase (with a high degree 
of accuracy equal to 0 in SM) and \rd, \ord\ are given by
\begin{equation}
\rd = \left| \frac{\langle \bar{K} \pi | \mathcal{H} \Dna}{ \langle K \pi| \mathcal{H} \Dna} \right|^2,
\qquad  \qquad
\ord = \left| \frac{\langle K \pi | \mathcal{H} \Dba}{ \langle \bar{K} \pi| \mathcal{H} \Dba} \right|^2.
\end{equation}
These ratios are proportional to $\left|V_{cd} V_{us}^{*} / V_{cd} V_{ud}^{*}
\right|^2 \sim \mathcal{O}(\tan^4 \theta_{c})$.
	 
To obtain the true values of the mixing parameters \textit{x, y}, knowledge of 
amplitude ratios \rd\ and strong phase difference \dkp\ is crucial. While 
$\left|\rd\right|$ can be determined from the fit to the $t$-dependent rate 
(Eq.~\ref{eq:tdepend}), it is not possible to extract the exact value
of the strong phase without supplemental studies. One method is to measure 
$\delta_{K\pi}$ using decays of coherent $\Dn-\Db$ pairs produced from the 
$\psi(3770)$ decay. This way $\cos{\delta}$ could be extracted from the 
interference between the $K^- \pi^+$ and CP eigenstate. The CP eigenstate 
tags the $K^- \pi^+$ to be the eigenstate with an opposite eigenvalue which 
represents a linear combination of $\Dn$ and $\Db$. The resulting decay rate 
for the second $D$-meson is modulated by the relative strong phase between 
$\Dn \to K^- \pi^+$ and $\Db \to K^- \pi^+$. Up to now the only measurement of 
$\sin{\delta}$  has been performed by the CLEO-c collaboration
using $D \to \Ks \pi^+ \pi^-$ as a tagging decay~\cite{Asner:2012xb}. 
However, this measurement suffers from the overall sign ambiguity that 
could not be resolved without external inputs and also leads to non-Gaussian 
uncertainties of $\delta$.

The method proposed in this paper allows one to measure the strong phase 
difference between DCS and CF decays with better accuracy and both 
$\cos{\delta}$ and $\sin{\delta}$ are measured simultaneously, thus achieving
uniform sensitivity in the whole $\delta$ range and resolving the ambiguity.

\section{${\mathbf {K^0}}$ evolution} \label{sec:k_evol}

The time evolution of the \Kn--\Kb\ system is described by the Schr\"odinger 
equation:
\begin{equation}
i {\partial}_t
{\Kn(t) 
\choose \Kb(t)}  = 
\left({{\mathbf{M}} - \frac{i}{2} {\mathbf {\Gamma}}}\right) 
{\Kn(t)   \choose \Kb(t)} \, ,
\label{eqn:schro}
\end{equation}
where the ${\mathbf M}$ and ${\mathbf \Gamma}$ matrices are Hermitian, 
and $CPT$ invariance requires $M_{11} = M_{22} \equiv M$ and $\Gamma_{11} = \Gamma_{22} \equiv \Gamma$. 

The Hamiltonian eigenvalues could be written as follows:
\beqn
m_{1,\,2}-i\frac{\Gamma_{1,\,2}}{2} = \left( M_{11} - i \frac{\Gamma_{11}}{2} \right) \pm \pqk
\left( M_{12} - i \frac{\Gamma_{12}}{2}  \right) \, , 
\eeqn 
where $m_{1,\,2}$ are masses, $\Gamma_{1,\,2}$ are widths of the Hamiltonian
eigenstates and parameters $p$, $q$ which correspond to the flavor admixtures 
of flavor states are defined by
\beq
\pqk^2 = \frac{M_{12}-\frac{i}{2}\Gamma_{12}}{M_{12}^*-\frac{i}{2} \Gamma_{12}^*} \, .
\eeq

For \Kn\ from the studied decays the following boundary conditions are met:
\beq
{\Kn(t)\choose \Kb(t)} \bigg|_{t=0}= {a \choose b} \, ,
\eeq
where $a=1$, $b=\sqrt{\rd} e^{i\delta}$ ignoring \ddb\ mixing which introduces 
just a tiny bias. 

From Eq.~(\ref{eqn:schro}) one can obtain the time evolution of the \Kn-meson 
produced as a linear combination $\al \Kna + \be \Kba$ into \Kn\ and \Kb:
\beqn
| \Kn(t) \rangle = \frac{1}{2} e^{-imt} e^{-\frac{1}{2} \Gamma t} \left[ a \left( 1 + e^{-i \Delta m t} e^{\frac{1}{2} \Delta \Gamma t} \right) +
b \pqk \left( 1-e^{-i \Delta m t} e^{\frac{1}{2} \Delta \Gamma t} \right) \right] , \nonumber \\
| \Kb(t) \rangle = \frac{1}{2} e^{-imt} e^{-\frac{1}{2} \Gamma t}  \left[ b \left( 1+e^{-i \Delta m t} e^{\frac{1}{2} \Delta \Gamma t} \right)  + a \qpk \left( 1 - e^{-i \Delta m t} e^{\frac{1}{2} \Delta \Gamma t} \right) \right] ,
\eeqn
where $m$ and $\Gamma$ are $\Ks$ mass and width respectively, $\Delta m = m_1 - m_2 > 0$, and $\Delta \Gamma = \Gamma_1 - \Gamma_2$. In this paper we consider only 
semileptonic final states and denote corresponding decay amplitudes as

\beq
A_{\lep^+} = \langle \pi^- \lep^+ \nu | \mathcal{H} \Kna , \qquad A_{\lep^-} = \langle \pi^+ \lep^- \bar{\nu} | \mathcal{H} \Kba
\eeq
The corresponding decay rates could be expressed as
\beqn
N_{\lep^+}(t)= \frac{1}{4} e^{-\Gamma t} |A_{\lep^+}|^2 \left[\left|a\right|^2 K_+(t) + \left|b\pqk\right|^2 K_-(t) + 2Re \left\{ ab \pqk K_i(t) \right\} \right] , \nonumber \\
N_{\lep^-}(t)= \frac{1}{4} e^{-\Gamma t} |A_{\lep^-}|^2 \left[ \left|a\right|^2 K_-(t) + \left|b\qpk \right|^2 K_+(t) + 2Re \left \{ ab \qpk K_i(t) \right\} \right],
\eeqn
where $K_{\pm, \,i}(t)$ are defined as
\beqn
K_{\pm}(t) = 1 \pm 2 e^{\frac{1}{2} \Delta \Gamma t} \cos(\Delta m t) + e^{\Delta \Gamma t}, \quad K_i(t) = 1 + 2 i e^{\frac{1}{2} \Delta \Gamma t} \sin(\Delta m t) - e^{\Delta \Gamma t}.
\eeqn

\begin{figure}[h]
\centering
\includegraphics[width=0.8\linewidth]{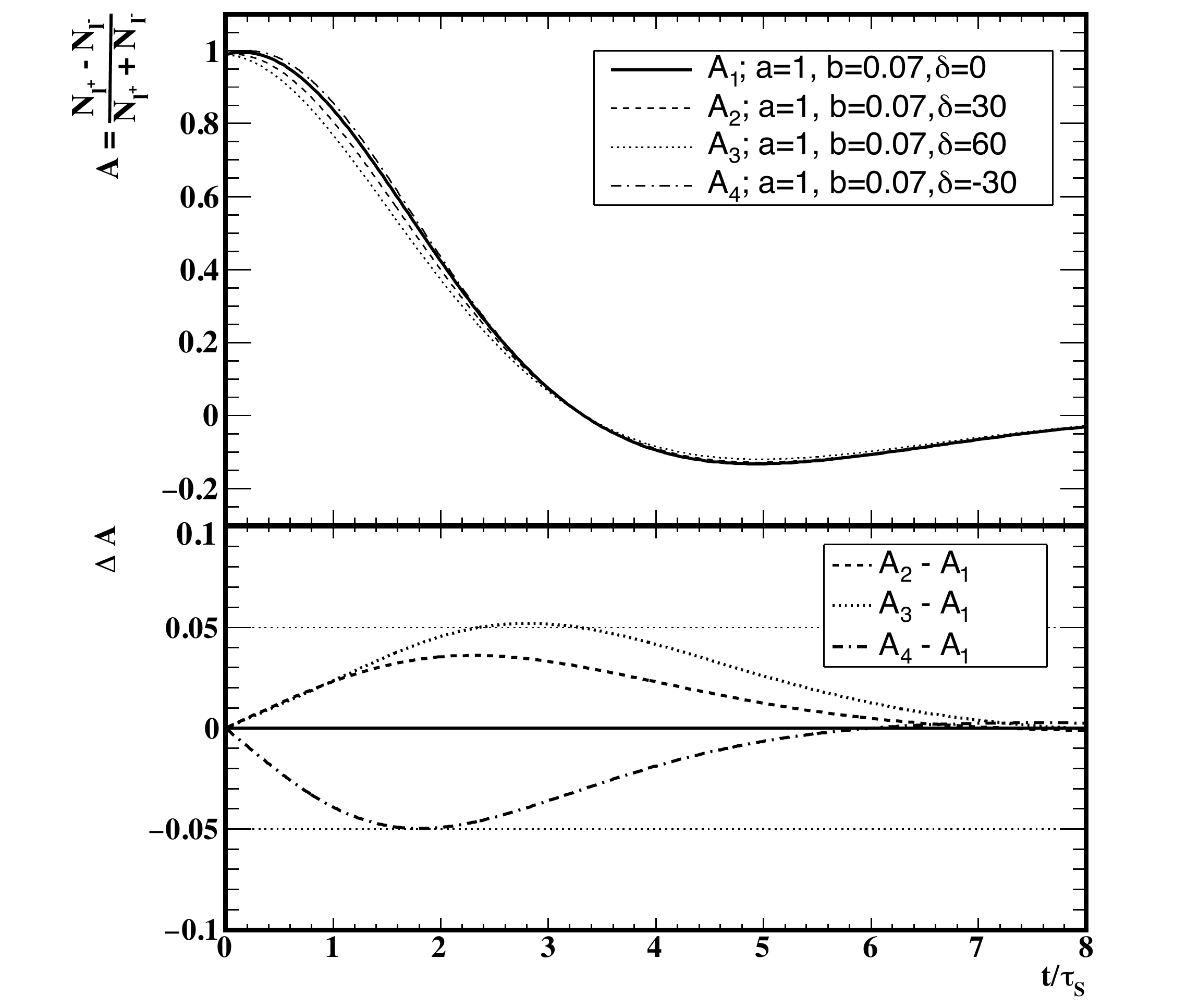}
\caption{{\it (top)} Lines show the lepton charge asymmetry for different 
initial admixtures of $a \Kna + be^{i\delta} \Kba$. The solid line corresponds 
to the case of exact $SU(3)$ symmetry and so $\delta = 0$. The dashed lines 
illustrate the obtained asymmetry for the values of the strong phase 
30$^{\circ}$, 90$^{\circ}$, 180$^{\circ}$;
{\it (bottom)} Here the dashed lines show the differences between asymmetries 
and $\delta=0$ case.} \label{Asy}
\end{figure}

The strong-phase difference $\delta$ enters the last term of each rate, and 
thus can be extracted from the measured $N_{\lep^\pm}(t)$. To illustrate effects 
induced by $\delta$ better, we form an "asymmetry" from these decay rates. 
Assuming that we have successfully tagged the flavor of \Dn\ in its decay time 
({\it {e. g.}} by a charge of a slow pion from the $D^{*+} \to D^0 \pi^+$)
decay, we are able to use the lepton charge to form the asymmetry in the 
following way:
\beq
\mathcal{A}(t) \equiv \frac{N_{\lep^+}(t) - N_{\lep^-}(t)}{N_{\lep^+}(t) + N_{\lep^-}(t)}.
\eeq
These asymmetries with different parameters $a$ and $b$ are shown in 
Fig.~\ref{Asy}. From this figure it can be seen that the most sensitive 
interval for the strong-phase $\delta$ lies in the range 
$\sim[0.5, 7]$ \Ks\ lifetimes.

\section{Feasibility study}\label{sec:feasib}
In this section we test the proposed method, obtain its potential efficiency 
and compare the accuracy with results from the CLEO-c~\cite{Asner:2012xb} and 
BESIII~\cite{Ablikim:2014gvw} experiments. This method could be applied 
in the LHCb and  Belle II experiments or at the future \ctau\ factory. 

\subsection{Reconstruction}
In order to perform time-dependent analysis, proper kaon vertex and momentum 
reconstruction is required. Because of the missing neutrino in the final 
state the direct kaon-momentum reconstruction is not possible. However, the 
kaon flight direction can be obtained from the primary and secondary vertices, 
while the momentum magnitude can be extracted from the measured pion and 
lepton momenta relying on the four-momentum conservation 
($(P_{K} - P_{\pi \lep})^2 = P_\nu^2 =0$) by constructing the following equation:
\beqn
\label{eq:4mom}
m_{K}^2 - 2\Epl \sqrt{{\bf p}_{K}^2 + m_{K}^2} + 2|{\bf p}_{K}| |\Ppl| \cos{\theta} + m_{\pi \lep}^2=0 \, ,
\eeqn 
where ${\bf p}_{K}$, $m_{K}$ are kaon three-momentum and mass, 
\Epl, \Ppl, $m_{\pi \lep}$ are energy, three-momentum and mass of the 
reconstructed pion and lepton combination, and $\theta$ is the angle between 
the \Kn\ direction obtained from the vertex information and measured momentum 
of the $\pi\lep$ combination. The magnitude of the kaon momentum, 
$|{\bf p}_{K}|$, is the only unknown, while all other terms in the equation 
are measured. Solving the quadratic equation~(\ref{eq:quad}) one can obtain 
two solutions for the kaon momentum
\beq
|{\bf p}_K|_{(1,2)} = \frac{|p_{\pi \lep}| \cos{\theta}(m_K^2 + m_{\pi l}^2) \pm \sqrt{w}}{2(\Eps - p_{\pi l}^2 \cos^2{\theta})},
\label{eq:quad}
\eeq
where $w$ is given by
\beq
w = \Eps \left( 4 m_K^2 \Pps \cos^2{\theta} - 4 \Eps m_K^2 + m_{\pi l}^2(m_K^2 + m_{\pi l}^2) \right).
\eeq

\subsection{Feasibility}

While the method can be applied in all high-luminosity experiments, in this 
section we consider only Belle II. At LHCb a measurement in the channel 
$\Dn \to \Kn \pin$ is somewhat problematic due to huge background in the 
neutral mode, however, one could select the kinematic region in 
$\Dn \to \Kn \pin$ decay at the expense of losing statistics, where \pin\ is 
energetic while \Kn\ is relatively soft. Both factors are favorable for this 
study, as a requirement of energetic \pin\ results in much smaller background, 
while soft \Kn\ has a smaller boost, thus its decay length is smaller, 
which increases the range of the accepted \Kn\ proper time in the 
relatively short LHCb tracking system. For estimates of the LHCb potential 
the toy Monte Carlo is not sufficient. For the charm-tau factory a high 
data sample for $D^{*+}\to\Dn\pi^+$ is anticipated, as part of the data will 
be taken above the $D^{*\pm}D^{\mp}$ threshold. However, a huge data sample is 
expected at the $\psi(3770)$ resonance. If tagging is performed using 
semileptonic decays of the second $D$-meson in the event, the situation is
equivalent to the $D^{*+}$ tagging. The interesting effects arise in case of 
hadronic tagging due to quantum correlations. A brief discussion of this case 
and its features could be found in Section~\ref{sec:appA}. 

Good tracking performance provides a sufficient vertex resolution 
($\sim \! 100 \, \mu$m) at Belle II, that results in the angular resolution 
of $\sim \! 2\,$mrad for a kaon with $t\gtrsim\tau_{K_{S}}$ and typical 
$\beta\gamma \sim 2$. In $\sim 35\%$ cases the discriminant is negative 
due to detector smearing; setting $w\equiv0$ in this case does not lead to 
degradation of the $|{\bf p}_{K}|$ resolution but eliminates the ambiguity. 
In $\sim30\%$ of positive $w$, when two-fold ambiguity arises, only one 
solution is physical, while the second one could be rejected because of 
the negative value obtained for the magnitude of the kaon momentum or 
the \Dn\ momentum exceeding a kinematic limit. In the remaining cases 
the correct solution can be selected by choosing those giving the \Dn\ mass 
closer to the expected one. 

\begin{figure}[h]
    \centering
    \includegraphics[width=\linewidth]{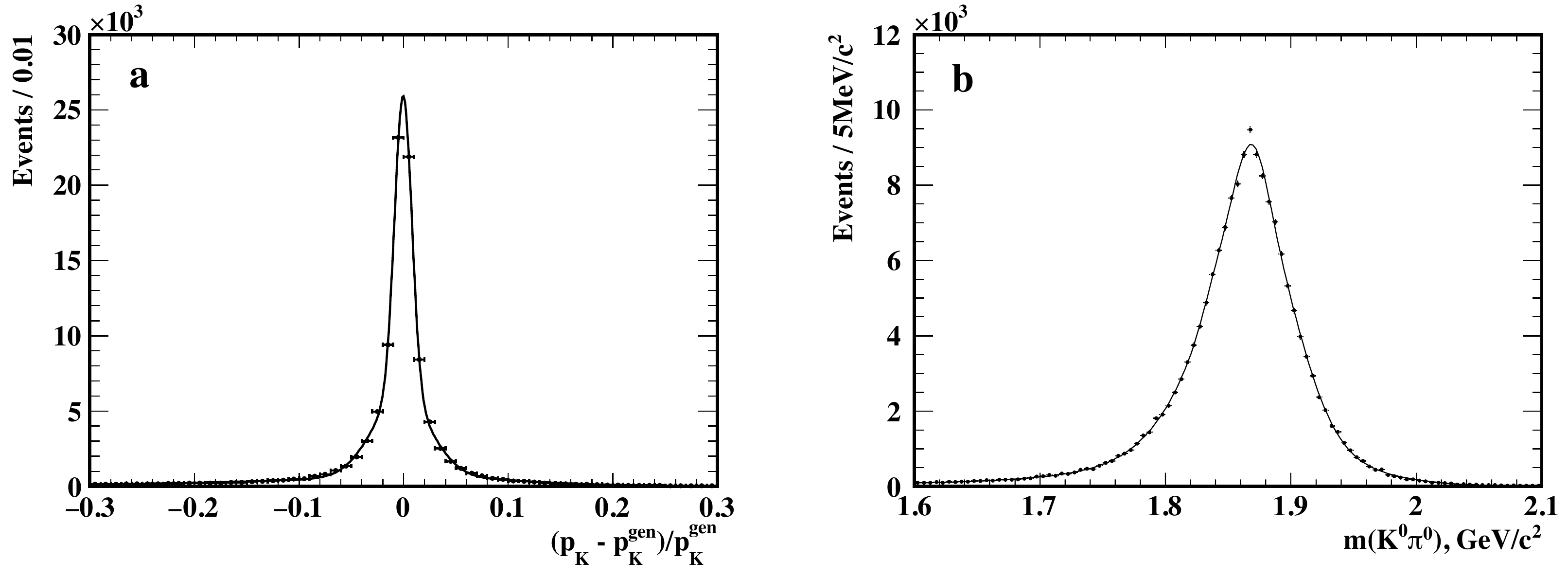}
    \caption{(\figa) Relative \Kn\ momentum resolution, (\figb)
    distribution of the $\Kn \pin$ invariant mass for selected candidates.}
    \label{d0_rec}
\end{figure}

For the selected candidates the resolution of \Kn\ momenta is shown in 
Fig.~\ref{d0_rec}~\figa\ as estimated from the toy Monte Carlo (MC) 
simulation with the actual detector tracking performance~\cite{Abe:2010gxa}. 
The resolution is estimated to be $\sigma_{p_{K}}/p_{K} \sim 0.02$. The 
$\Dn\to\Kn\pin$ mass resolution ($\sim 40 \mevc$) is dominated by \pin, 
rather than \Kn\ momentum resolution (Fig.~\ref{d0_rec}~\figb). 

\begin{figure}[h]
    \centering
    \includegraphics[width=\linewidth]{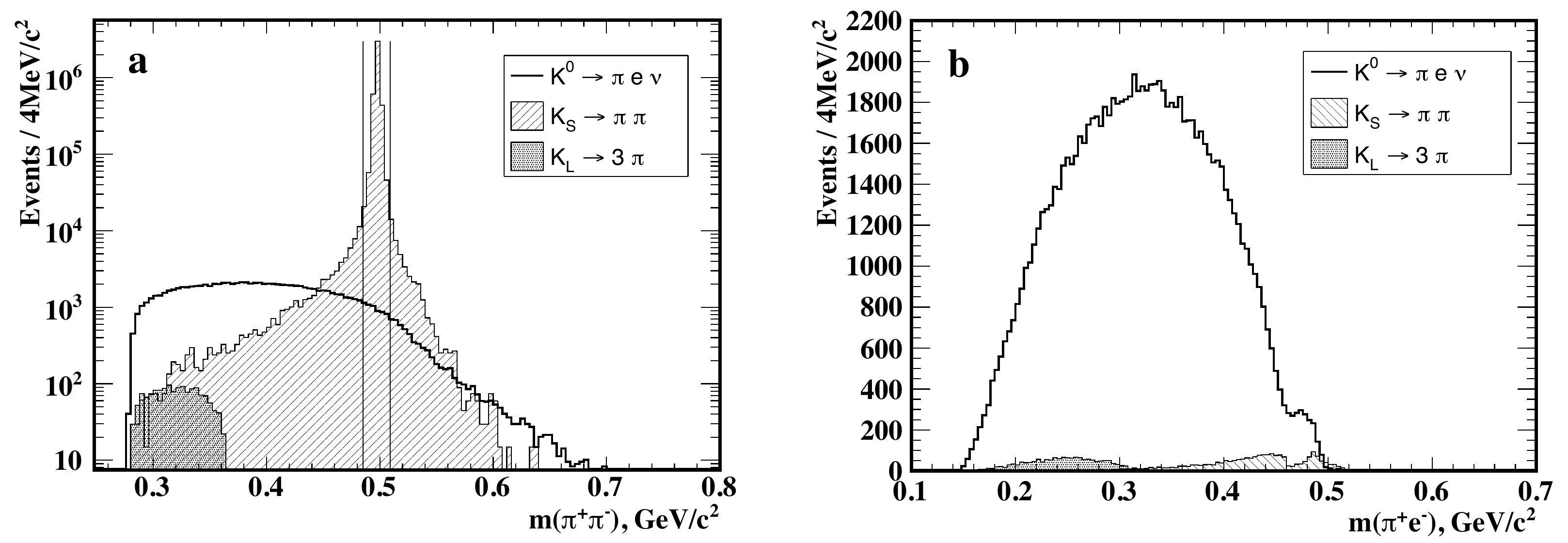}
    \caption{(\figa) Distribution of the $\pi^+ \pi^-$ invariant mass. 
The contributions from three processes are drawn. Here the pion mass is 
also assigned to an electron from $K^0 \rightarrow \pi e \nu_{e}$ decay. 
The vertical lines show the proposed $\Ks$ veto. Applying this veto one 
can obtain the $e \pi$ invariant mass distribution shown in the right 
(\figb) where hatched histograms represent the background contribution. }
    \label{fig:bgr}
\end{figure}

It is expected that backgrounds can be suppressed to a level at which they 
are dominated by combinations of real $\Kn \to \pi^\pm\lep^\mp \nu$ with \pin. 
Rejecting $\pi^\pm$ or $\lep^\mp$ coming from the primary vertex, suppresses 
fake secondary vertices formed by random intersection of primary tracks. 
While this requirement suppresses a signal with a short kaon lifetime, the 
affected part of the signal is not interesting for our study. The true 
secondary vertices can be produced by such kaon decays as $\Ks\to\pp$, 
$\Kl\to\pp\pin$, $K^+\to\pi^+\pp$, or by secondary interactions. 
$\Ks\to\pp$ is initially a huge background source that exceeds the signal 
by two orders of magnitude. However, it can be efficiently vetoed by 
requiring the mass of the $\pi^\pm\lep^\mp$ combination in the "pion" mass 
hypothesis (when the $\pi^{\pm}$ mass is ascribed to both tracks from the 
secondary vertex in a calculation of the mass of the combination) to be 
outside the nominal \Ks\ mass window. Figure~\ref{fig:bgr}~\figa\ shows 
the secondary vertex mass in the "pion" mass hypothesis for the signal and 
different backgrounds, where a huge $\Ks \to \pp$ signal can be easily rejected
at the expense of the $\sim\!(2-3)\%$ loss of the signal efficiency. 
After the \Ks\ veto all real strange vertices becomes smaller than the signal. 
The $\pi^\pm \lep^\mp$ mass spectrum remaining after the \Ks\ veto is shown in 
Fig.~\ref{fig:bgr}~\figb. Lepton identification suppresses the remaining 
non-$\Kn\to \pi^\pm\lep^\mp\nu$ background down to a negligible level, taking 
into account that the typical misidentification rate is $(0.5-2)\%$.

\begin{figure}[h]
\centering
\label{fig:pktk}
\includegraphics[width=0.6\linewidth]{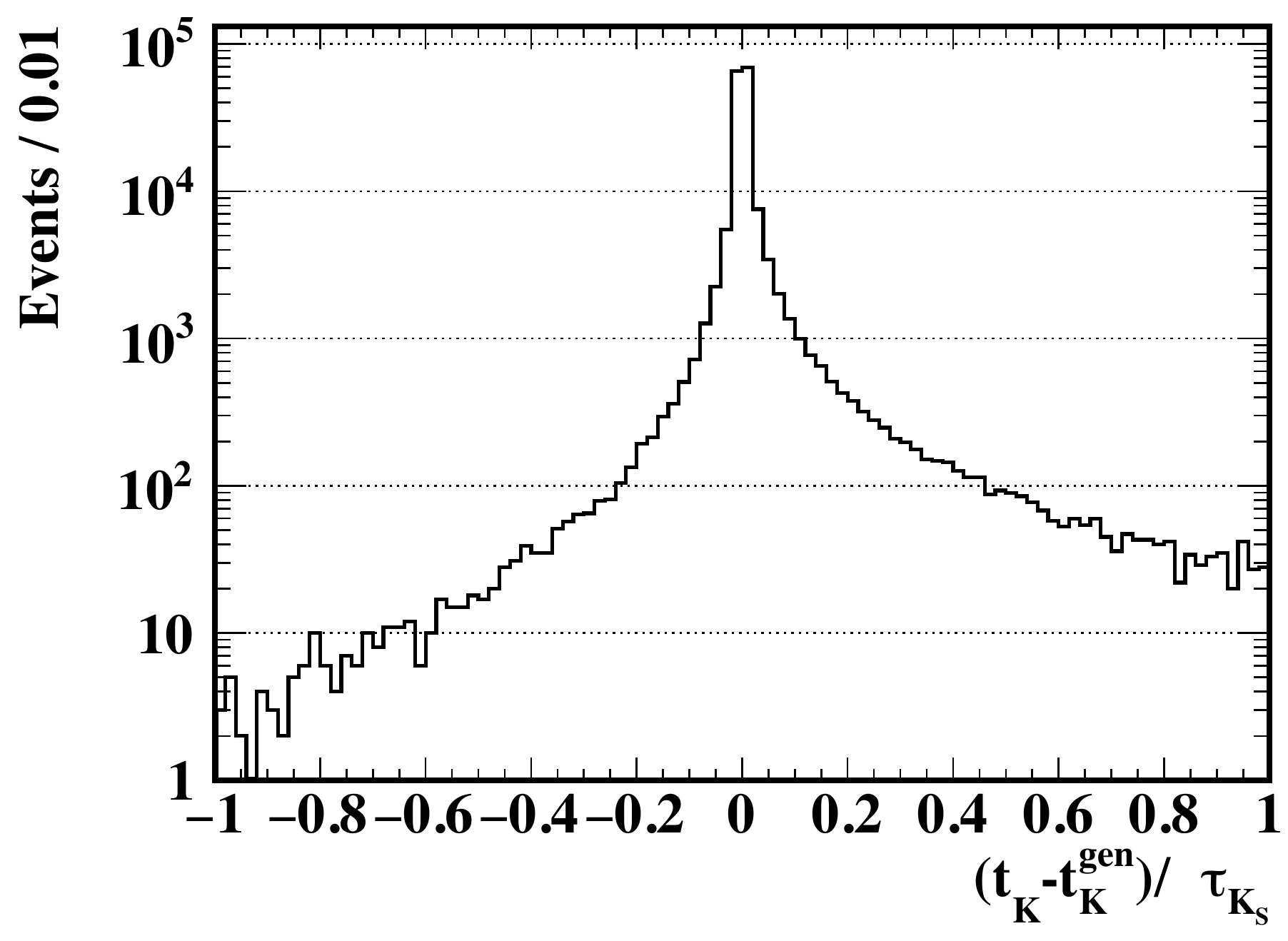}
\caption{Resulting kaon lifetime resolution.}
\end{figure}

To test potential accuracy of the proposed method and to confirm that there is 
no bias, we generate 200 MC samples of $10^5$ events, each with a value of 
the angle $\delta$ in the $[-90^{\circ}, 90^{\circ}]$ interval with a step of 
$10^{\circ}$. An estimation of the approximate number of events is based on 
a full data sample that will be collected in the Belle II experiment 
($50\mathrm{ab}^{-1}$), corresponding branching fractions ($10^{-5}$) and 
reconstruction efficiency ($\sim 20\%$). In order to extract $\delta$ we 
perform a simultaneous fit to both "right-" and "wrong-sign'' histograms. We 
limit the fit range to the [$0.5$, $8$] of kaon lifetimes where the highest 
sensitivity could be achieved.

\begin{figure}[h]
    \centering
    \includegraphics[width=0.75\linewidth]{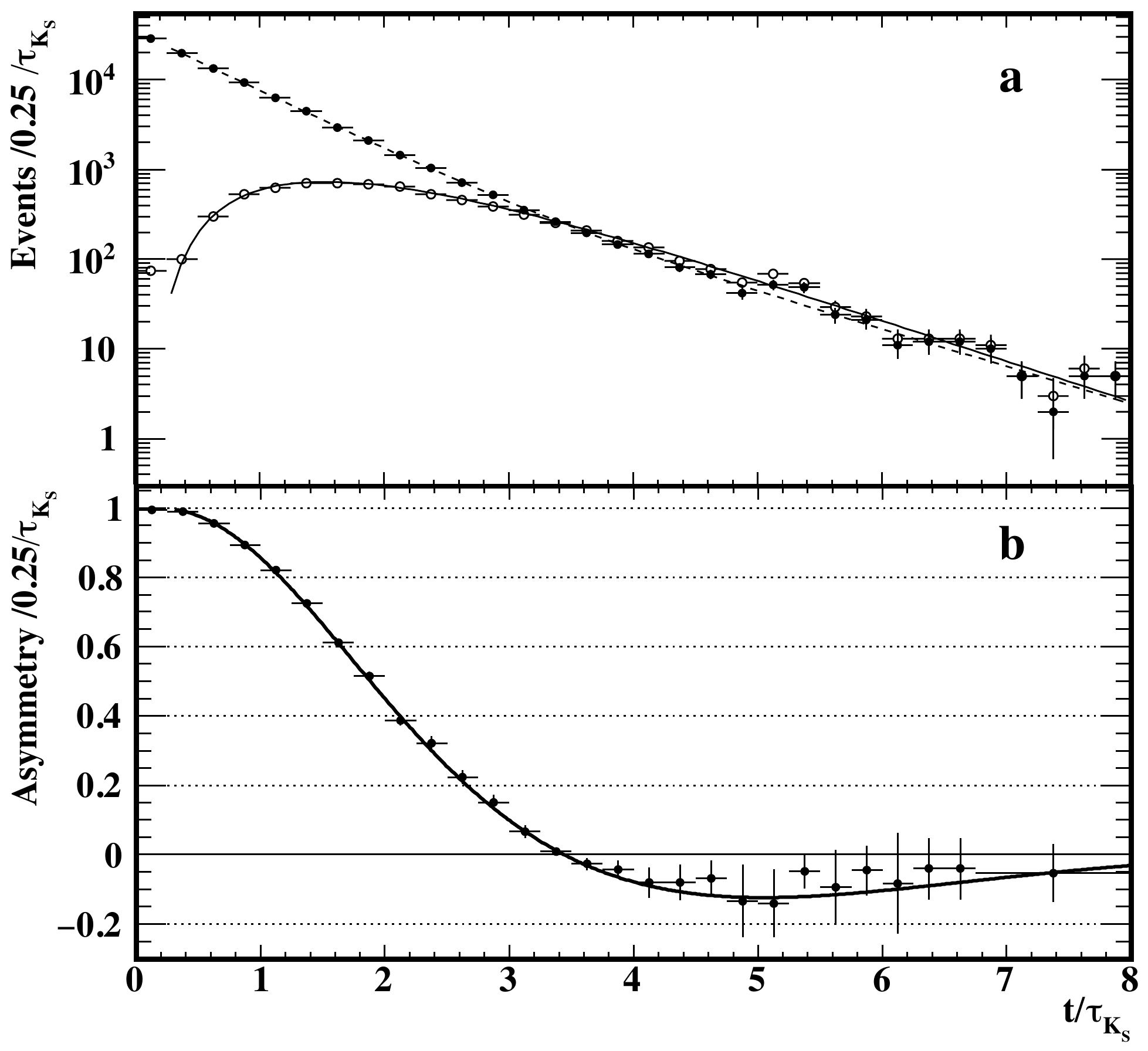}
    \caption{(\figa) "Toy" MC generated time-dependent decay rates. The 
histogram with filled black circles corresponds to "right sign" and 
with open circles correspond to the "wrong sign" decay rates. The solid and 
dashed lines correspond to the result of the simultaneous fit. (\figb) 
The resulting decay rate asymmetry with the fit result superimposed.}
    \label{fig:mainRes}
\end{figure}

\begin{figure}[h]
    \centering
    \includegraphics[width=\linewidth]{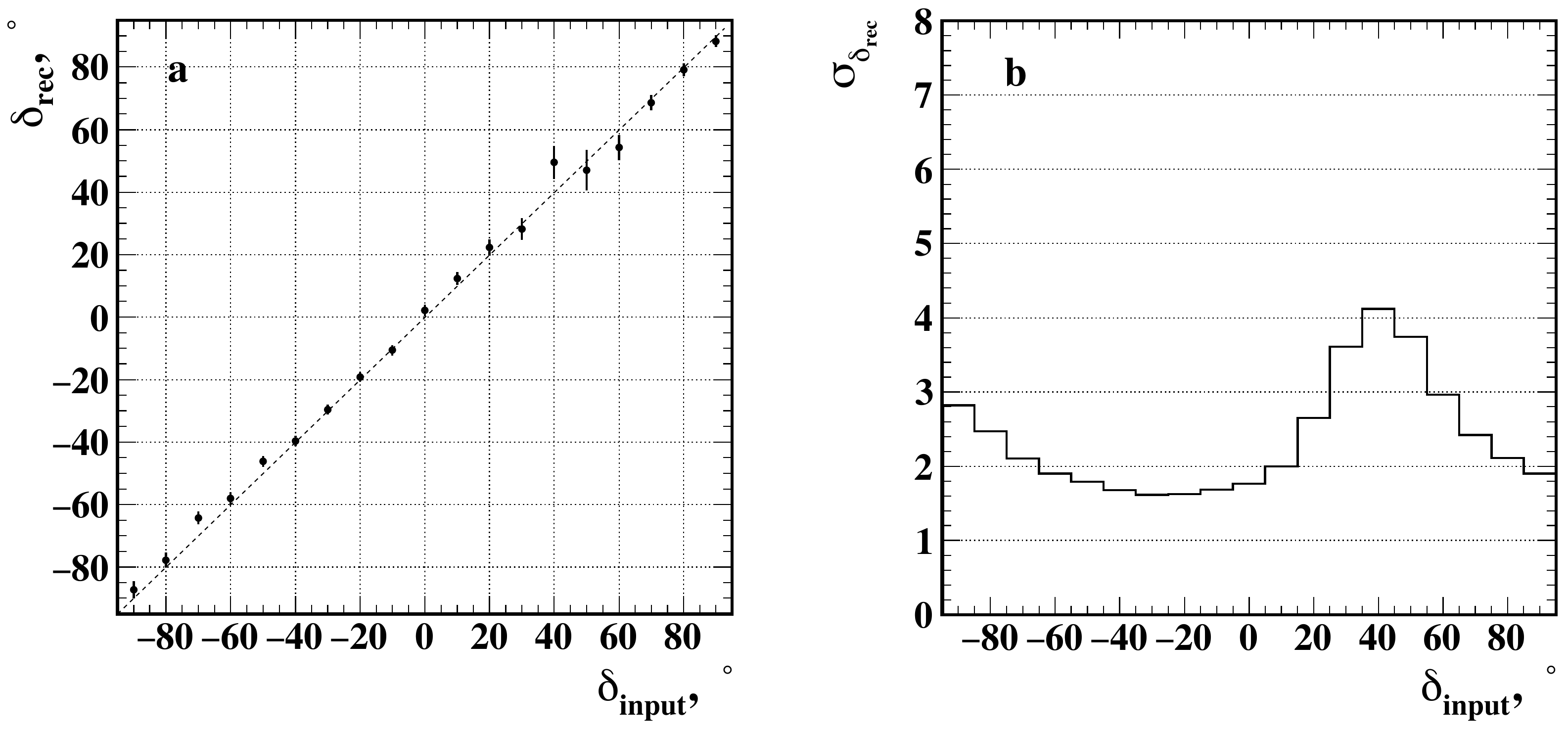}
    \caption{(\figa) Resulting $\delta$ values extracted from the fit 
with the obtained uncertainty (\figb).}
    \label{fig:feasRes}
\end{figure}

Each pair of histograms is fitted 20 times with different initial values 
of the $\delta$ parameter to ensure that a fit is converged to a global 
rather than a local minimum independently of the starting value. The value 
obtained with the best $\chi^2$ is chosen. 
Figure~\ref{fig:mainRes}~\figa\ illustrates the fit results for a MC sample 
with $\delta = 20^{\circ}$ which is close to the central value published by 
PDG~\cite{Tanabashi:2018oca}, and Fig.~\ref{fig:mainRes}~\figb\ shows the 
asymmetry with the fitted asymmetry superimposed. The results for the 
whole range of generated $\delta$ values is illustrated in 
Fig.~\ref{fig:feasRes}~\figa.

The resulting distribution for the uncertainty is shown in 
Fig.~\ref{fig:feasRes}~\figb. In the region of small values of the strong 
phase (which are consistent with the previous measurements of CLEO-c and 
BESIII) the obtained uncertainty basically is under $4^{\circ}$. This fact makes the proposed method comparable to the measurement at BESIII with a $20\,\mathrm{fb}^{-1}$ data sample at the $\psi(3770)$ proposed in Ref.~\cite{Cheng:2007uj}. 

The procedure mentioned was performed with 200 MC samples. The resulting 
distribution of obtained values for the strong phase was fitted and the 
Gaussian mean and standard deviation were extracted. As an example, we 
present here the obtained distribution for $\delta = 20^{\circ}$. The extracted 
mean and standard deviation are consistent with results of  individual fits. 

\begin{figure}[h]
    \centering
    \includegraphics[width=\linewidth]{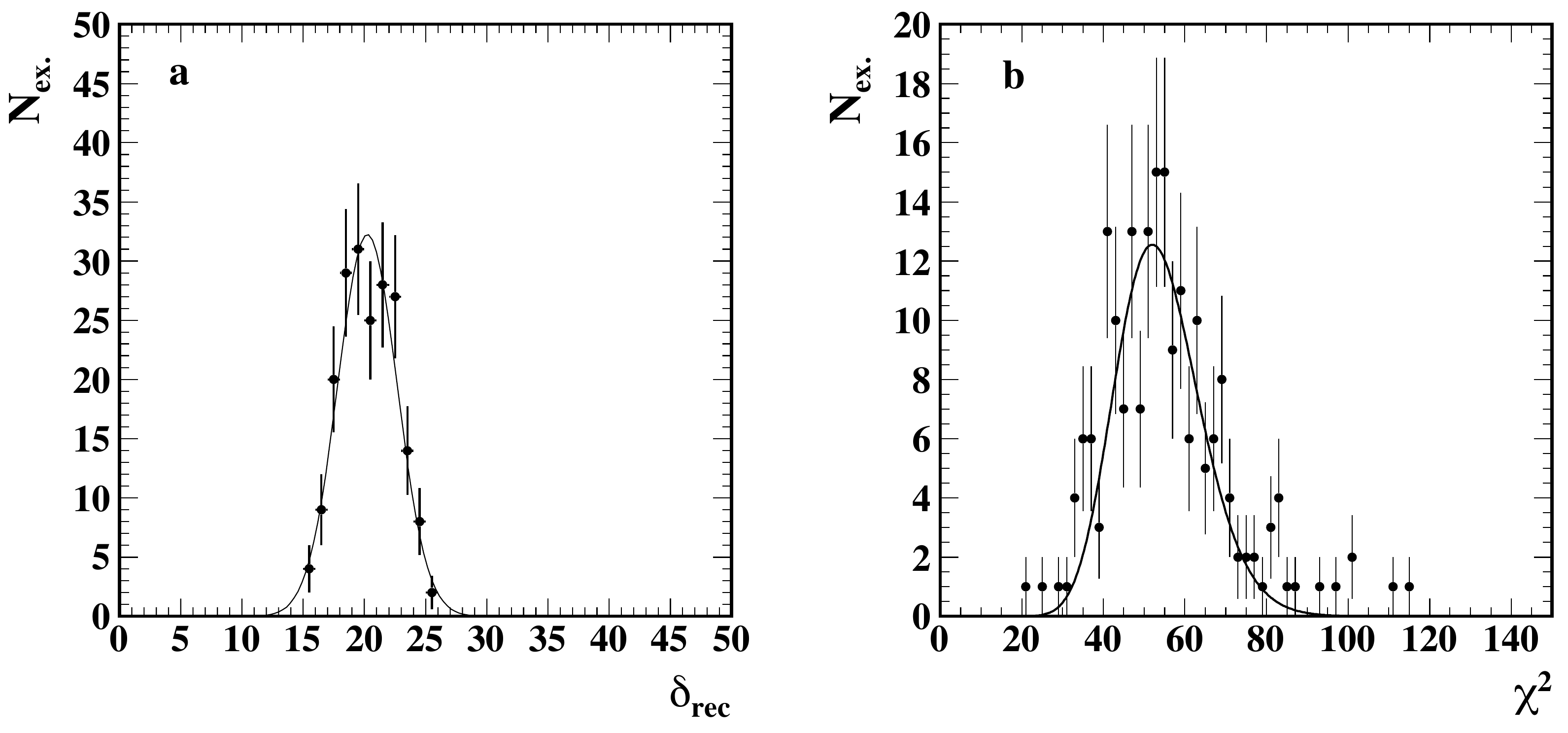}
    \caption{ (\figa) Distribution of $\delta$ values obtained in 200 
pseudoexperiments and (\figb) distribution of chi-square values. The solid 
lines are fit results with Gaussian and chi-square distribution functions,
respectively.}
\end{figure}

\section{Conclusion}
In this paper we have performed a phenomenological study of the evolution 
of the neutral kaons produced in the decay $\Dn \to \Kb \pi^0$. Usage of a 
semileptonic final state allows us to tag the kaon-flavor final state, 
hence providing sensitivity to a strong-phase measurement. As it was shown 
in Section \ref{sec:k_evol}, both $\sin{\delta}$ and $\cos{\delta}$ contribute 
to the resulting decay rate. This way it is possible to measure $\delta$ 
without trigonometrical ambiguity. 
It was shown that the effect induced by the strong phase in the decay rate 
asymmetry depends on the $\delta$ value. Fortunately, the interval most 
sensitive to the measurement allows us to effectively cut off the background 
coming from the interaction region. Using MC simulation background produced 
by true secondary vertexes has been studied. In Section \ref{sec:feasib} 
it was shown that these types of background could be suppressed down to 
a negligible level.

To test the proposed method, a feasibility study was performed. The full data 
sample of $50 ab^{-1}$ that will be collected in the Belle II experiment was 
used as an approximate amount of data. The feasibility study showed that 
potential accuracy of the proposed method is comparable to the classic 
method based on the $\psi(3770) \to \Dn \Db$ study \cite{Cheng:2007uj}.

\acknowledgments
We wish to thank Simon Eidelman for useful discussions.
The reported study of V.P. was funded by RFBR, project number 19-32-90104.
Work of P.P. is supported by the Russian Ministry of Science and Higher Education contract 14.W03.31.0026.

\section{Appendix A: Correlated $\mathbf{\Dn \Db}$ pair case}
\label{sec:appA}

For the future \ctau\ factory it seems more reasonable to use correlated 
$D$-mesons from the $\psi(3770) \to \Dn \Db$ decay, since the \ctau\ factory 
will be able to produce $\sim 2 \times 10^9$ $\psi(3770)$ mesons per 
year~\cite{Bondar:2013cja, Chen:1997at}. The pair of $D$-mesons is produced in the $1^{--}$ 
state and is described by the following wave function
\beq
\Psi _{D\bar{D}} = \frac{1}{\sqrt{2}} [ |\Dn_{phys}(t)\rangle |\Db_{phys}(t)\rangle -  |\Db_{phys}(t)\rangle |\Dn_{phys}(t)\rangle],
\eeq
where $|\Dn_{phys}(t)\rangle$, $|\Db_{phys}(t)\rangle$ describe time evolution 
of initially pure states defined in (\ref{eq:devol}). The time-dependent 
decay rate for the given system could be expressed as \cite{Xing:1994mn} 
\begin{multline}
    R(f_1, t_1, f_2, t_2) \propto |A_{f_1}|^2 |A_{f_2}|^2 e^{- \Gamma (t_1+t_2)} \Big[ \frac{1}{2} |\xi + \zeta|^2 e^{- \Delta \Gamma/2 (t_2-t_1)}
    + \frac{1}{2} |\xi - \zeta|^2 e^{\Delta \Gamma/2 (t_2-t_1)} - \\ - (|\xi|^2 - |\zeta|^2) \cos{(\Delta m (t_2-t_1))} + 2 Im(\xi^{*} \zeta) \sin{(\Delta m (t_2-t_1))} \Big],
    \label{eq:corrd}
\end{multline}
where $A_{f_1}$, $A_{f_2}$ -- amplitudes of the $D$-meson decay to the final 
states $f_1$ and $f_2$, respevtively, $t_1$ and $t_2$ -- proper decay times 
for $D$-mesons, $\Delta m$ and $\Delta \Gamma$ -- mass and width difference 
for states $D_1$ and $D_2$, and parameters $\xi$ and $\zeta$ are given by 
\beq
\xi = \left( \frac{p}{q} \right)_D - \left( \frac{q}{p} \right)_D \frac{\overline{A}_{f_1}}{A_{f_1}} \frac{\overline{A}_{f_2}}{A_{f_2}},
\eeq
\beq
\zeta = \frac{\overline{A}_{f_2}}{A_{f_2}} - \frac{\overline{A}_{f_1}}{A_{f_1}},
\eeq
where
\beq
A_{f_i}=\langle f_i |H| \Dn \rangle, \, \, \overline A_{f_i} = \langle f_i |H| \Db \rangle. 
\eeq
Eq.~\ref{eq:corrd} could be also written in the following way:
\begin{multline}
     R(f_1, t_1, f_2, t_2) \propto |A_{f_1}|^2 |A_{f_2}|^2 e^{- \Gamma (t_1+t_2)} \Big[ (|\xi|^2 + |\zeta|^2) \cosh{(\Delta \Gamma \Delta t/2)} + \\ + 2 Re(\xi \zeta)\sinh{(\Delta \Gamma \Delta t/2)} -
     (|\xi|^2 - |\zeta|^2) \cos{(\Delta m \Delta t)} + 2 Im(\xi^{*} \zeta) \sin{(\Delta m \Delta t)} \Big],
\end{multline}
If we assume here that $\Delta \Gamma \ll 1$ and $\Delta m \ll 1$, 
the decay rate could be expressed in terms of $x$ and $y$ as follows:
\begin{multline}
    R(f_1, t_1, f_2, t_2) \propto |A_{f_1}|^2 |A_{f_2}|^2 e^{- \Gamma (t_1+t_2)} \Big[ 2 |\zeta|^2 + \frac{1}{2} |\xi|^2 \left(\Gamma \Delta t \right)^2 (x^2+y^2) +  \\ + \frac{1}{2} |\zeta|^2 \left(\Gamma \Delta t \right)^2 (y^2-x^2) + 2 \Gamma \Delta t \left\{ y Re(\xi \zeta)  + x Im(\xi^* \zeta) \right\} \Big] , 
\end{multline}

An extensive study of different combinations of $D$-meson decay can be found 
in Ref.~\cite{Gronau:2001nr}. However, the decay modes like 
$D \rightarrow \Kn \pi^0$ were not considered there. Here we consider the 
following combinations of the $ \big\{ K^- \pi^+; \Kb \pi^0 \big\}$ and $\left\{X \lep^- \nu ; \Kn \pi^0\right\}$ final states:
\begin{itemize}
    \item[(1)] The final state $ \big\{ K^- \pi^+; \Kb \pi^0 \big\}$ is 
particularly interesting because both strong phases $\dzz$ and $\dpm$ are 
present in the resulting amplitude. Parameters $\xi$ and $\zeta$ will 
be given by 
        \beq
        \xi = \pqd -  \qpd \sqrt{\rdz} \sqrt{\rdp} e^{i (\dpm + \dzz)},
        \eeq
        \beq
        \zeta = \sqrt{\rdz} e^{i \dzz} - \sqrt{\rdp e^{i \dpm}},
        \eeq
    And the expression for the decay rate is
    \begin{multline}
        R(t_1,t_2) \propto 2|A_{K^- \pi^+}|^2 |A_{\Kb \pi^0}|^2 e^{- \Gamma (t_1+t_2)} \Biggl[
        \left( \rdz + \rdp - 2 \sqrt{\rdz \rdp} \cos{(\dzz - \dpm)} \right) + \\
        + \left( \left|\frac{p}{q}\right|^2_D + \left|\frac{q}{p}\right|^2_D \rdz \rdp - 2\sqrt{\rdz} \sqrt{\rdp} \cos{(\dzz + \dpm)} \right) \frac{( \Gamma \Delta t )^2}{4} (x^2+y^2) + \\
        + \Gamma \Delta t \Delta_{12} \Biggr],
        \label{eq:kpkz}
    \end{multline}
    where we have neglected the term $\propto (y^2-x^2)$ and defined a new 
variable  
    \beq{}
    \Delta_{12} = \sqrt{\rdp} \left( \pqd x' + \qpd \rdz y' \right) -
    \sqrt{\rdz} \left( \pqd x'' + \pqd \rdp y'' \right).
    \eeq
    Comparing the given result to the amplitude of the final state 
$ \big\{ K^- \pi^+; K^- \pi^+ \big\}$ which is accessible only through mixing,
one can notice that a new component not proportional to mixing arises. 
Relying on the isospin $SU(2)$ symmetry we assumed previously that 
$\dzz \simeq \dpm$. The amplitude ratios $\rdz$, $\rdp$, however, could not 
be assumed equal even in the strict $SU(2)$ limit. A difference in internal 
and external $W$-boson emission as well as color suppression implies that 
$\rdz \neq \rdp $. The latter guarantees that the resulting decay rate 
will have a non-vanishing component independent of $\ddb$ mixing. 
    
    \item[(2)] A semileptonic final state for one of the $D$-mesons implies 
that $\overline{A}_{f_1} /A_{f_1} = 0$. In such a case we could expect 
the amplitude to be similar to the case 
$\left\{X \lep^- \nu ; K^+ \pi^- \right\}$ with substitution 
$\rdp \to \rdz$ and $\dpm \to \dzz$. In this case one can obtain
    \beq
        \xi = \left( \frac{p}{q} \right)_D,
    \eeq
    \beq
        \zeta = \sqrt{\rdz} e^{i \dpm}
    \eeq
    
    \begin{multline}
        R(t_1,t_2) \propto 2|A_{X l \nu }|^2 |A_{\Kb \pi^0}|^2 e^{- \Gamma (t_1+t_2)} \Biggl[ \rdz + \left|\frac{p}{q}\right|^2_D \frac{( \Gamma \Delta t )^2}{4} (x^2+y^2) + \\ + \pqd \Gamma \Delta t \sqrt{\rdz} y'' \Biggr] , 
    \end{multline}

\end{itemize}

\begin{thebibliography}{99}

\bibitem{Isidori:2010kg}
  G.~Isidori, Y.~Nir, and G.~Perez,
  ``Flavor Physics Constraints for Physics Beyond the Standard Model,''
  Ann.\ Rev.\ Nucl.\ Part.\ Sci.\  {\bf 60}, 355 (2010).

\bibitem{Aubert:2007wf}
  B.~Aubert {\it et al.} (BaBar Collaboration),
  ``Evidence for \ddb\ mixing'',
  Phys.\ Rev.\ Lett.\  {\bf 98},  211802 (2007).
 
 \bibitem{Staric:2007dt}
  M.~Staric {\it et al.} (BELLE Collaboration),
 ``Evidence for \ddb\ mixing'',
  Phys.\ Rev.\ Lett.\  {\bf 98}, 211803 (2007). 

\bibitem{Aaij:2019kcg}
  R.~Aaij {\it et al.} [LHCb Collaboration],
  ``Observation of CP Violation in Charm Decays,''
  Phys.\ Rev.\ Lett.\  {\bf 122},  211803 (2019).
  
\bibitem{Bazavov:2017weg}
  A.~Bazavov {\it et al.},
  ``Short-distance matrix elements for \Dn-meson mixing for $N_f=2+1$ lattice QCD,''
  Phys.\ Rev.\ D {\bf 97},  034513 (2018). 

\bibitem{Chang:2017zaw}
  C.~C.~Chang {\it et al.} [Fermilab Lattice and MILC Collaborations],
 ``$D$-Meson Mixing in 2+1-Flavor Lattice QCD,''
  PoS LATTICE {\bf 2016}, 307 (2017). 

\bibitem{Simone:2016bbg}
  J.~Simone,
  ``Neutral $B$-meson and $D$-meson mixing matrix elements from $2+1$ flavor lattice QCD,''
  PoS LATTICE {\bf 2015}, 332  (2016).

\bibitem{Ko:2014qvu}
  B.~R.~Ko {\it et al.} [Belle Collaboration],
  ``Observation of \ddb\ Mixing in \ee\ Collisions,''
  Phys.\ Rev.\ Lett.\  {\bf 112}, 111801 (2014),  
   Addendum: [ibid.  139903].
  
\bibitem{CIBINETTO:2014jsa}
  G.~Cibinetto [BaBar Collaboration],
  ``Charm Mixing and CP Violation at B-Factories,''
  Int.\ J.\ Mod.\ Phys.\ Conf.\ Ser.\  {\bf 35}, 1460413 (2014).
  
\bibitem{Aaij:2016roz}
  R.~Aaij {\it et al.} (LHCb Collaboration),
 ``Measurements of charm mixing and CP violation using $\Dn \to K^\pm \pi^\mp$ decays'',
  Phys.\ Rev.\ D {\bf 95}, 052004 (2017), 
   Erratum: Phys.\ Rev.\ D {\bf 96}, 099907 (2017).

\bibitem{Amhis:2016xyh}
  Y.~Amhis {\it et al.} [HFLAV Collaboration],
  ``Averages of $b$-hadron, $c$-hadron, and $\tau$-lepton properties as of summer 2016,''
  Eur.\ Phys.\ J.\ C {\bf 77},  895 (2017). 

\bibitem{Bediaga:2018lhg}
 R.~Aaij {\it et al.} [LHCb Collaboration],
  ``Physics case for an LHCb Upgrade II - Opportunities in flavour physics, and beyond, in the HL-LHC era,''
  {\tt arXiv:1808.08865} (2018).
  
\bibitem{Abe:2010gxa} 
  T.~Abe {\it et al.} [Belle-II Collaboration],
  ``Belle II Technical Design Report'',
 {\tt arXiv:1011.0352} (2010).

\bibitem{Chen:1997at}
  H.~S.~Chen, ``Tau charm factory project at Beijing,''
  Nucl. Phys. Proc. Suppl.  {\bf 59}, 316 (1997).
  
\bibitem{Bondar:2013cja}
  A.~E.~Bondar {\it et al.} [Charm-Tau Factory Collaboration],
 ``Project of a Super Charm-Tau factory at the Budker Institute of Nuclear Physics in Novosibirsk,'' Phys. Atom. Nucl. {\bf 76}, 1072 (2013).
 
 \bibitem{Bergmann:2000id}
  S.~Bergmann, Y.~Grossman, Z.~Ligeti, Y.~Nir, and A.~A.~Petrov,
  ``Lessons from CLEO and FOCUS measurements of \ddb\ mixing parameters'',
  Phys. Lett. B {\bf 486}, 418 (2000). 

\bibitem{Asner:2012xb} 
  D.~M.~Asner {\it et al.} [CLEO Collaboration],
  ``Updated Measurement of the Strong Phase in $\Dn \to K^+\pi^-$ Decay Using Quantum Correlations in $\ee \to \Dn\Db$ at CLEO,''
  Phys. Rev. D {\bf 86}, 112001 (2012).
  
\bibitem{Ablikim:2014gvw} 
  M.~Ablikim {\it et al.} [BESIII Collaboration],
  ``Measurement of the $\Dn \to K^-\pi^+$ strong phase difference in $\psi(3770)\to \Dn\Db$,''
  Phys.\ Lett.\ B {\bf 734}, 227 (2014).
  
\bibitem{Tanabashi:2018oca} 
  M.~Tanabashi {\it et al.} [Particle Data Group],
  ``Review of Particle Physics,''
  Phys.\ Rev.\ D {\bf 98}, 030001 (2018).

\bibitem{Cheng:2007uj}
  X.~D.~Cheng {\it et al.},
  ``Strong phase and \ddb\  mixing at BES-III'',
  Phys.\ Rev.\ D {\bf 75}, 094019  (2007).

\bibitem{Xing:1994mn} 
  Z.~z.~Xing,
  ``Time dependence of coherent $P^0 \, \bar{P}^0$ decays and CP violation at asymmetric B factories,''
  Phys.\ Rev.\ D {\bf 53}, 204 (1996).
 
\bibitem{Gronau:2001nr} 
  M.~Gronau, Y.~Grossman, and J.~L.~Rosner,
  ``Measuring \ddb\ mixing and relative strong phases at a charm factory,''
  Phys.\ Lett.\ B {\bf 508}, 37 (2001).


\end{thebibliography}
\end{document}